\documentstyle[preprint,aps]{revtex}
\begin{document}
\title{Magnetic enhancement of Co$_{0.2}$Zn$_{0.8}$Fe$_2$O$_4$ spinel oxide by 
mechanical milling}
\author{R. N. Bhowmik\footnote{e-mail:rnb@cmp.saha.ernet.in} and R.
Ranganathan\footnote{e-mail:ranga@cmp.saha.ernet.in}}
\address{Experimental Condensed Matter Physics Division,\\ 
Saha Institute of Nuclear Physics, 1/AF, Bidhannagar, Calcutta 700064, India}
\author{ S. Sarkar, C. Bansal}
\address{ School of Physics, Central University of Hyderabad, India} 
\author{R. Nagarajan}
\address{Tata Institute of Fundamental Research, India}
\maketitle
\begin{abstract}

We report the magnetic properties of mechanically milled  
Co$_{0.2}$Zn$_{0.8}$Fe$_2$O$_4$ spinel oxide. 
After 24 hours milling of the bulk sample, the XRD spectra show
nanostructure with average particle size $\approx$ 20 nm. 
The as milled sample shows an enhancement in magnetization and ordering 
temperature compared to the bulk sample. 
If the as milled
sample is annealed at different temperatures for the same duration, 
recrystallization process occurs and approaches to the bulk structure on
increasing the annealing temperatures. 
The magnetization of the annealed samples first increases
and then decreases. At higher 
annealing temperature ($\sim$ 1000$^{0}$C) the system shows two coexisting 
magnetic phases {\it i.e.}, spin glass state and ferrimagnetic state, 
similar to the as prepared bulk sample. 
The room temperature M\"{o}ssbauer spectra of the as milled 
sample, annealed at 300$^{0}$C for different durations (upto 575 hours), 
suggest that the observed change in magnetic behaviour
is strongly related with cations redistribution between tetrahedral (A) and
octahedral (O) sites in the spinel structure.
Apart from the cation redistribution, we suggest that the enhancement of 
magnetization and ordering temperature is related with the reduction of B site 
spin canting and increase of strain induced anisotropic energy during 
mechanical milling.

\end{abstract}

\section{Introduction}
In recent years, several research groups \cite{Fatemi,Soren,Goya,Fine} are 
involved in the investigations of nanoparticle spinel oxides because of their 
potential applications in magnetic devices, in micro wave technology \cite{Buschow}, 
in high density magnetic recording media \cite{Kyder}, in magnetic fluids  
as drug carrier etc. \cite{Fine,Hefi}. 
Various types of nanoparticle materials such as, metal: Fe, Co, Ni \cite{Khan}, 
metallic alloys:  Fe-Cu \cite{Crespo}, and metallic 
oxides: MnFe$_2$O$_4$ \cite{Soren} and ZnFe$_2$O$_4$ \cite{Hamdeh}, 
are under current research activity. 
While metal and inter metallic nanoparticles suffer from stability problems
in atmospheric condition, metallic oxides are highly stable under ambient
conditions \cite{Chen}. Various factors such as, particle size
distribution \cite{Fine}, inter-particle interactions \cite{Fiorani}, 
grain (core) and grain boundary (shell) structure
 \cite{Bianco,Kodama} and metastable structure of the system \cite{Grig}
control the properties of nanoparticles. Some of the specific 
properties of the nanoparticles which are of interest are Quantum magnetic tunneling \cite{Fine},
various magnetic order like ferrimagnet/ferromagnet, spin
glass/superparamagnet and spin canting effects \cite{Kodama,Zeng}.\\ 
The interesting aspect of magnetism in spinel oxides is that the magnetic 
order is strongly dependent 
on the competition between various superexchange type interactions i.e., 
J$_{AB}$ (A-O-B) and J$_{BB}$ (B-O-B), where A: tetrahedral (A) sites moments
and B: octahedral (B) sites moments, O: is O$^{2-}$ ions \cite{Buschow}. 
Certain amount of site disorder {\it i.e.}, the cations redistribution between A and 
B sites is sufficient to change the super-exchange interactions in
nanoparticles spinel \cite{Braber,Zaag}. Hence, the magnetic properties of 
the nanoparticle spinels can be drastically different from their bulk
counterpart. In this light, we investigated Co$_{0.2}$Zn$_{0.8}$Fe$_2$O$_4$
spinel oxide system.\\ 
The cation distribution of the bulk sample is
(Zn$^{2+}_{0.8}$Fe$^{+3}_{0.2}$)$_A$[Co$^{+2}_{0.2}$Fe$^{3+}_{1.8}$]$_B$O$_4$
( A= tetrahedral sites, B= octahedral sites) \cite{Petitt,rnb}.
The A site is strongly diluted with respect to B site and  
B site moments contribute to different size clusters. The 
spins inside the clusters form canted structure due to short range
ferrimagnetic interactions. The total magnetization M (=
M$_B$cos$\theta$-M$_A$) of such type of canted system depends not only on the
sublattice (site) magnetization M$_B$ and M$_A$ but also on the canting angle
$\theta$ between the B site spins. The B site spin canting is strongly related with the
amount of A site moments. If more number of A site moments exist in 
as milled Co$_{0.2}$Zn$_{0.8}$Fe$_2$O$_4$ sample, the canting angle between B
site spins is expected to be reduced and a drastic change in cluster glass 
properties of bulk sample \cite{rnb} is expected. 
Unfortunately, most of the reports in literature concern only on the long range order
system, such as ZnFe$_2$O$_4$ \cite{Hamdeh} and NiFe$_2$O$_4$ \cite{Kodama}, and
the magnitudes of A and B
site moments but not on the short range interacting cluster glass system where 
change of spin canting inside the clusters plays an important role 
 \cite{Hamdeh}.  
In this presentation, we show that for as milled Co$_{0.2}$Zn$_{0.8}$Fe$_2$O$_4$ sample, the
magnetization enhancement is related with the reduction of B site spin canting
and an enhancement of ordering temperature occurs which is related with the 
cation redistribution and strain induced anisotropy.

\section{Experimental}

\subsection{Sample preparation}

Stoichiometric amounts of ZnO (99.998\% from Johnson Matthey),
Fe$_2$O$_3$ (99.998\% from Johnson Matthey) and Co$_3$O$_4$ (99.5\% from Fluka) 
oxides for the composition Co$_{0.2}$Zn$_{0.8}$Fe$_2$O$_4$ were mixed and
ground for 2 hours. Then, the mixture was pelletized and sintered at 950$^0$C
for 12 hours and slowly cooled to room temperature. The pellet was again
ground and pelletized and finally sintered at 1000$^0$C for 12 hours. The
heat treatment was carried out under atmospheric condition with heating and 
cooling rate of 2-3 $^0$C/minute. The XRD spectra of the as prepared bulk 
sample (S$_0$) confirm formation of well crystalline cubic spinel phase.
The as prepared sample was milled 
for 24 hours in a SPEX 8000 mixer/mill using a set of six balls, two of
diameter 1/2 inch and four of diameter 1/4 inch with 
ball to powder weight ratio 5:1.
The as milled sample (S$_{1}$) was heat treated in air at 300$^{0}$C for various 
durations, ranging from 2 hours to 575 hours, to study the temporal phase
evolution. The 6 hours heat treated as milled samples are denoted as
S$_{N3}$ (for 300$^0$C), S$_{N6}$ (for 600$^{0}$C) and S$_{N10}$ (for 1000$^{0}$C). The heating and cooling rates were 
maintained at 2-3$^0$C/minute during the heat treatment process. 

\subsection{X-ray diffraction}

The X ray diffraction (XRD) data (Fig. 1a) were taken using Philips PW1710 
diffractometer 
with CuK$_{\alpha}$ radiation in the 2$\theta$ range 10$^{0}$-90$^{0}$ with a 
step size 0.02$^0$. The XRD data of as prepared bulk sample 
show narrow crystalline lines with estimated particle size of a few
micrometers. After 24 hours milling of the bulk sample, the
XRD peak lines become significantly broad and no other phases are seen, except
the cubic spinel structure. 
The broadening in XRD lines suggest nano structure 
or non-uniform micro strain introduced in the lattices during milling process
 \cite{Cill}. 
The width of $<311>$ XRD peak line gives average particle size $\approx$ 20 nm 
for the as milled sample (S$_{1}$). 
When the as milled sample is heat treated, the XRD lines (Fig.1b-c) become 
sharp, which indicates recrystallization process in the system. 
The average particle size increases with annealing temperature
as 23 nm, 28 nm and 62 nm for the samples S$_{N3}$, S$_{N6}$ and S$_{N10}$,
respectively. Eventhough the XRD spectrum of the S$_{N10}$ 
sample is very similar to that of S$_0$ sample, the average particle size
$\approx$ 62 nm suggest that complete 
restoration of crystallinity close to S$_0$ sample is yet to be reached. 

\subsection{X-ray Fluorescence spectra}

We have determined the chemical composition of S$_{1}$ sample by X-ray 
Fluorescence (XRF) technique. 
The Fluorescence spectrum was detected 
using ORTEC HPGe detector and finally recorded in a PC based multichhanel
analyzer. The XRF spectrum (Fig. 2) shows the characteristic K$_{\alpha}$ and
K$_{\beta}$ lines of Fe, Co and Zn elements. No extra lines have been found
that correspond contamination from hardened steel balls with composition 
Fe$_{74}$Cr$_{18}$Ni$_{8}$. The peak integrals of the K$_{\alpha}$
lines of all the elements are obtained using the standard peak fitting
programme to get the relative percentage. This program uses the
principle of Fundamental Parameter Method \cite{Jenkin}. The obtained
composition is : Fe = 62.45\%, Co = 7.17\% and Zn = 30.38\% which is very close
to the expected values {\it i.e.} Fe = 63.53\%, Co = 6.71\% and Zn = 29.76\%
for Co$_{0.2}$Zn$_{0.8}$Fe$_2$O$_4$ system. The maximum error of the fitted 
values are within 5\%.

\subsection{Measurements}

The low field dc magnetization (T = 20K to 320K, H= 10 Oe to
100 Oe) and ac susceptibility (T= 60K to 330K, h$_{rms}$ $\approx$ 1 Oe and
frequency (f) = 337 Hz and 7 kHz) measurements have been
performed using home made magnetometer \cite{Anindita}. 
In ZFC condition, the sample has been cooled from 
room temperature to 20K in the absence of dc magnetic field, then the field was 
applied at 20K and magnetization
data were recorded while increasing the temperature.  
In FC condition, the sample was cooled from
room temperature to 20K in presence of dc magnetic field (same magnitude
which was applied during ZFC measurement) and the magnetization data were 
recorded with increasing temperature and keeping the field on.  
High field magnetization, hysteresis 
experiments were performed using VSM magnetometer in the fields 
upto $\pm$12 Tesla in the temperature range 10K to 300K.\\
M\"{o}ssbauer spectra were recorded at 300K without applying external magnetic
field, using a constant acceleration spectrometer in transmission geometry
mode. The spectra were recorded using a
5 mCi $^{57}$Co in Rh matrix source. The hyperfine magnetic field (HMF)
distribution at the $^{57}$Fe nuclei was evaluated from the M\"{o}ssbauer
spectra using the method of Le Ca\"{e}r and Dubois \cite{Cear}. In this model
a linear relationship between HMF (H) and isomer shift (IS) is assumed in the 
form IS = {\it a}H + {\it b}, where {\it a} and {\it b} are the fitting 
parameters to get a minimum $\chi^{2}$. The isomer shift (IS)  
wes calculated with respect to $^{57}$Fe metal.

\section{Results and Discussion}

\subsection{AC susceptibility}

Fig. 3 shows the real ($\chi^\prime$) and imaginary ($\chi^{\prime\prime}$)
components of ac susceptibility for sample S$_{1}$.  
It shows a broad $\chi^\prime$ maximum at T$_B$ $\approx$
320K (for h$_{rms}$ $\approx$ 1 Oe, f= 337 Hz) and the corresponding 
$\chi^{\prime\prime}$ maximum at
 $\approx$ 290K. The broadness of the ac susceptibility maxima suggest a 
cluster size distribution in the as milled sample S$_{1}$. 
The temperature where $\chi^\prime$ shows maximum is defined
as the average blocking temperature (T$_B$) of the clusters.
From the frequency dependence of ac susceptibility measurements (Fig. 3a), a 
positive frequency shift of T$_B$ is observed in the 
$\chi^\prime$  as well as in the $\chi^{\prime\prime}$ maximum
in the measurement frequency range. This
is a characteristic of superparamagnetic blocking of the ferromagnetic 
clusters in different metastable states \cite{Hamdeh}.
In the low temperature region, the $\chi^\prime$ data show a small 
shoulder at $\sim$ 100K (designated as T$_K$) and $\chi^{\prime\prime}$ data 
show (Fig. 3b) increasing tendency below T$_K$. 
This type of behaviour of T$_K$ can be attributed due to the
disordered surface or grain boundary spins in a nanoparticle system \cite{Kodama}.\\
In order to check the effect of heat treatment on magnetic properties, we have 
carried out the ac susceptibility measurements for the S$_{N10}$ sample. 
This heat treated sample shows mixed magnetic phases. The $\chi^\prime$ and
$\chi^{\prime\prime}$ data of S$_{N10}$ sample (Fig. 4) show two magnetic
ordering at T$_{m1}$ $\approx$ 70K and T$_{m2}$ $\approx$ 280K respectively.
The frequency shift of T$_f$ follows Vogel-Fulcher law 
\begin{equation}
f = f_{0}exp^{-E_{a}/(T_{m1}-T_{0})}
\end{equation}
with characteristic frequency $f_{0}$ $\approx$ 10$^{10}$ Hz, activation
energy E$_a$ $\approx$ 379 eV and constant T$_0$ $\approx$ 47K. The
corresponding frequency shift per decade of {\it f} ($\Delta
T_{f}$/T$_{f}{\Delta}log(f)$) is $\approx$ 0.05, which characterizes 
spin glass like ordering at T$_{m1}$ \cite{Mydosh}. The ordering temperature
at T$_{m2}$ does not show significant frequency shift (not shown in Fig. 4), 
suggest ferrimagnetic ordering temperature at T$_{m2}$ $\approx$ 280K. 
The spin glass transition temperature 
T$_{m1}$ $\approx$ 70K and ferrimagnetic ordering temperature at T$_{m2}$
$\approx$ 280K of S$_{N10}$ 
sample are slightly different with respect to the
cluster spin glass freezing temperature T$_{m1}$ $\approx$ 110K and short range
ferrimagnetic ordering temperature T$_{m}$ $\approx$ 260K for bulk sample
 \cite{rnb}. But it is clear that the mixture of two 
magnetic phases (spin glass/cluster spin glass plus ferrimagnetic phase) of
S$_{N10}$ sample is very similar to the bulk sample S$_0$.
\subsection{DC magnetization}

The ZFC magnetization of S$_{1}$ sample (Fig. 5) shows a broad maximum at
the cluster blocking temperature T$_B$ $\approx$ 320K (H $\sim$ 30 Oe) with a 
thermomagnetic irreversibility
between ZFC and FC magnetization below T$_B$.
The decrease of ZFC
magnetization below T$_B$ is due to the blocking of ferromagnetic clusters
in different metastable states, whereas the FC magnetization increases below
T$_B$ due to
the orientation of these ferromagnetic clusters in metastable states 
which give rise to more magnetic contribution \cite{Chen}. 
It is observed (Fig.5a and Fig. 5b) that the T$_B$ is highly applied field 
dependent ({\it i.e.}, T$_B$ $\approx$ 320K, 280K and 50K for applied field 30 Oe, 
100 Oe and 1 Tesla, respectively) as expected for superparamagnetic
blocking of the clusters \cite{Chen}. 
The superparamagnetic behaviour of the as milled sample (S$_{1}$) can be 
further identified from the temperature dependence of field cooled 
thermoremanent 
magnetization (TRM) data (Fig. 5b inset) that  
the TRM value reduces to zero at T$\approx$ T$_B$. 
The zero TRM value indicate that
the effective inter-cluster interactions are also negligible and the 
clusters behave as a non-interacting small particle above T$_B$ $\approx$ 
320K.\\ 
An interesting feature is observed when we compare the magnetization data at 
100 Oe of as prepared bulk sample, as 
milled sample and heat treated samples (Fig. 6). 
The magnetization is enhanced when the bulk sample is
milled for 24 hours. On heat treating the as milled sample, it is
observed that magnetization further increases for S$_{N3}$ sample.
Then magnetization decreases for the S$_{N6}$ sample. If the as milled sample
is heat treated at higher temperature (S$_{N10}$), the magnetization
increases. But compared to the bulk sample S$_0$, S$_{N10}$ sample has a
higher magnetization value at low temperature and lower value at temperatures
$>$ 200K. Similar non-equilibrium 
magnetic behaviour as a function of annealing temperature 
has been observed for NiFe$_2$O$_4$ \cite{Berk}. 
The superparamagnetic blocking temperature T$_{B}$ (indicated in Fig. 6)
decreases to 270K and 120K for S$_{N3}$ and S$_{N6}$, respectively, in
comparison with T$_{B}$ $\approx$ 280K for S$_{1}$ sample at H= 100 Oe. 
The 100 Oe magnetization data of the S$_{N10}$ sample indicate a mixture of
two magnetic ordering at T$_{m1}$ $\approx$ 70K and T$_{m2}$ $\approx$ 225K,
respectively, very similar to the cluster spin freezing temperature at 
T$_{m1}$ $\approx$ 100K and short range ferrimagnetic ordering temperature at 
T$_{m2}$ $\approx$ 230K, respectively, for bulk sample \cite{rnb}. 
\subsection{Hysteresis}

The field dependence of magnetization data (Fig. 7) under ZFC  
condition is similar to that of ferromagnetic isotherms but the magnetization 
lacks saturation even upto 12 Tesla at any temperature. 
This, we attribute to the spin canting effects at grain 
boundary \cite{Kodama} or superparamagnetic contribution of nanoparticles
 \cite{Grig}. We note from Fig. 7a (inset) that the S$_{1}$ sample has a 
better ferromagnetic behaviour in terms of saturation with respect to magnetic field which indicates
that the strong spin canting behaviour of S$_0$ sample \cite{Petitt,rnb} has
decreased for the nanoparticle S$_{1}$ sample. Importantly, we also note
(Fig.7a, inset) that the saturation magnetization of nanoparticle sample
S$_{1}$ is significantly increased at room temperature with respect to the
bulk sample S$_{0}$. However, the higher value of M for S$_0$ at 10K for H $>$
4T compared to S$_{1}$ can be understood by assuming two types of magnetic
interactions in the system. One is ferromagnetic and second one is
antiferromagnetic. In S$_0$ sample the B site antiferromagnetic interactions 
are dominant which causes spin canting in B site and shows non-saturation in
magnetization \cite{rnb}. The reduction of antiferromagnetic interactions 
between the B site moments gives rise the 'better' ferromagnetic behaviour,
yet having a reduced saturation moment at 10K, for S$_{1}$ sample.
It is observed from the hysteresis loop (Fig. 7b) that 
isothermal remanent magnetization (M$_R$), 
coercive field (H$_C$) where M$_R$ is zero, and irreversible field
(H$_{irr}$), where the hysteresis loop closes,  
decrease with 
temperature. There is no hysteresis loop at T $\geq$ 290K. This suggest that
the thermal energy close to the blocking temperature ($\sim T_B$) 
is sufficient to reduce the coercive field to zero value \cite{Muri}.
The saturation magnetization (M$_S$) values, obtained from M vs 1/H plot,
are small in comparison with the bulk sample as seen in
nanocrystalline NiFe$_2$O$_4$ \cite{Berk}.
This is possible if the B site intra-cluster spin canting decreases or if
there is a redistribution of cations between A and B sites when the bulk
sample is mechanically milled \cite{Naran}. 

\subsection{M\"{o}ssbauer Spectroscopy}
Fig. 8 shows the room temperature M\"{o}ssbauer spectra in absence of any
external magnetic
field for the samples (S$_0$), S$_{1}$, S$_{N3}$, S$_{N6}$ and S$_{N10}$. 
The 
spectrum of S$_0$ sample consists of a Lorentzian doublet arising from the
Fe$^{3+}$ ions at octahedral (B) site and a Lorentzian singlet arising from
the Fe$^{3+}$ ions at tetrahedral (A) site. 
The most probable cations distribution of bulk sample (S$_0$) obtained is 
(Zn$_{0.8}$Fe$_{0.2}$)$_A$[Co$_{0.2}$Fe$_{1.8}$]$_B$O$_4$.
The isomer shift (IS) and quadrupole 
splitting (QS) of Fe$^{3+}$ ions at [B] site are 
$\approx$ +0.29 mm/sec and 0.35 mm/sec respectively. The IS  
value of Fe$^{3+}$ ions at (A) site is $\approx$ -0.056 mm/sec \cite{rnb}.
The M\"{o}ssbauer spectrum of the bulk sample therefore confirm that
Zn$^{2+}$ ions occupy A site, Co$^{2+}$ ions prefer B site and Fe$^{3+}$ ions
prefers both A and B sites. 
The M\"{o}ssbauer spectrum of the S$_{1}$ sample clearly indicates a hyperfine 
magnetic field splitting in addition to a central paramagnetic doublet. 
This spectrum of S$_{1}$ sample represents the appearance of spontaneous 
magnetization of ferromagnetic
clusters mixed with superparamagnetic fluctuation effect due to nano meter size
of the particles \cite{Hamdeh,Naran}. 
As the annealing temperature increases, the hyperfine magnetic field splitting
decreases for S$_{N3}$ sample and the S$_{N6}$
and S$_{N10}$ samples show paramagnetic spectra.
The M\"{o}ssbauer parameters of S$_{N10}$ sample (IS= +0.27 mm/sec and QS =
+0.43 mm/sec for B site Fe$^{3+}$ ions and the IS = + 0.25
mm/sec of A site Fe$^{3+}$ ions) differ from that of the as prepared bulk 
sample which suggests that duration of heat treatment ($\sim$ 6 hours) is not
enough to achieve 100\% equilibrium state of the bulk samples. 
These results confirm that the enhancement of magnetic
ordering and magnetization observed in S$_{1}$ sample is
intrinsic property of the material. 
These also further indirectly confirm the hypothesis of the presence of 
non-equilibrium cation distribution in the as milled sample.\\
To check the recovery of cation distribution from non-equilibrium 
state to equilibrium state, we recorded the M\"{o}ssbauer spectrum
of S$_{1}$ sample as function of duration of heat treatment at 300$^{0}$C.
Fig. 9 shows the M\"{o}ssbauer spectra and corresponding 
hyperfine magnetic field (HMF) distribution (p(H)).
The presence of multiple hyperfine fields at A and B site Fe$^{3+}$ ions can be
expected due to various super transferred hyperfine fields (STH) from
neighbouring ions.
The HMF acting on the B site Fe$^{3+}$ ions is due to non-uniform
environment, consisting of different number of Zn$^{2+}$ and Fe$^{3+}$ ions at 
the nearest neighbour A sites. The HMF experienced by the A site Fe$^{3+}$ 
ions is due to nearest neighbour B site configuration \cite{Petitt}. 
It is clear from the p(H) distribution that the changes
brought about by 2 hours heat treatment is not very significant but the peak 
intensities of hyperfine field components are decreasing and 
intensities of paramagnetic components are increasing for annealing time $\geq$
16 hours. Indirectly, it gives the information that
more and more number of Zn$^{2+}$ ions stabilizing in A site with the
compromise of more Fe$^{3+}$ ions in B site. This reduces the reduction of
super transferred hyperfine magnetic field via inter-sublattice superexchange
interactions \cite{Petitt}.
Fig. 10 shows the temporal evolution of the average HMF for 300$^{0}$C heat 
treated sample. 
The solid line represent the fit data using a function of time (t) as 
F(t) = {\it a} + {\it b}(1-exp(t/t$_0$)) where {\it a}, {\it b} and t$_0$ are 
constants. F(t) represents average HMF as a function of time. The fit gives a 
measure of
the reaction kinetics through atomic diffusion as the sample gets annealed.

\section{Summary and Conclusions}
Usually nano materials lead to the decrease 
in magnetization and ordering temperature as the particle size decreases 
 \cite{Chen,Soren}. This is already observed in  
Co$_{0.2}$Zn$_{0.8}$Fe$_2$O$_4$ with particle size $\sim$ 6nm to 70 nm, prepared
by coprecipitation method \cite{Sanjib}.
However, the bulk Co$_{0.2}$Zn$_{0.8}$Fe$_2$O$_4$ sample with short range
ferrimagnetic ordering temperature $\sim$ 260K and paramagnetic at 300K
 \cite{rnb} after 24 hours mechanical milling gives nanoparticles $\sim$ 20 nm
which shows ferromagnetic cluster blocking state below T$_B$ $\approx$ 320K 
with an enhancement of magnetization. 
Because of the weak inter-sublattice interactions J$_{AB}$ (due to low A site 
Fe$^{3+}$ population), the B site spins form canted structure
in bulk Co$_{0.2}$Zn$_{0.8}$Fe$_2$O$_4$ sample. 
The stable cation distribution of the bulk sample is destroyed due to 
mechanical milling and a non-equilibrium disordered state, {\it i.e.} the 
increase of A site Fe$^{3+}$ population enhances the inter-sublattice exchange 
interactions Fe$^{3+}_{A}$-O$^{2-}$-Fe$^{3+}_{B}$ (J$_{AB}$) and 
change the B sublattice superexchange interactions 
Fe$^{3+}_{B}$-O$^{2-}$-Fe$^{3+}_{B}$. 
This makes the clusters more ferromagnetic with lower spin canting.
The reduction of spin canting inside the clusters enhances the
magnetization. The superparamagnetic fluctuation effect is expected due to the
nano size clusters (particles). The cations redistribution surely is the main
factor which controls the magnetic properties of nanoparticle system but
distortion of spherical shape and strain induced anisotropy energy due to
mechanical milling can give rise to a preferential orientation of the spins
inside the clusters.\\
The magnetic reversibility of the heat treated (S$_{N10}$) milled sample 
towards the bulk sample (may not 100\%, due to
different micro structural parameters like particle size, cluster size 
distribution, spin canting angle inside the clusters) 
conclusively show that enhancement
of magnetization and ordering temperature of the as milled samples are intrinsic property of Co$_{0.2}$Zn$_{0.8}$Fe$_2$O$_4$ spinel oxide. 
\vspace{0.5 truecm}\\
\noindent Acknowledgement:
One of the authors RNB thanks Council of Scientific and Industrial Research 
(CSIR, India) for providing fellowship [F.No.9/489(30)/98-EMR-I ].

\newpage
\centerline{Figure Caption}
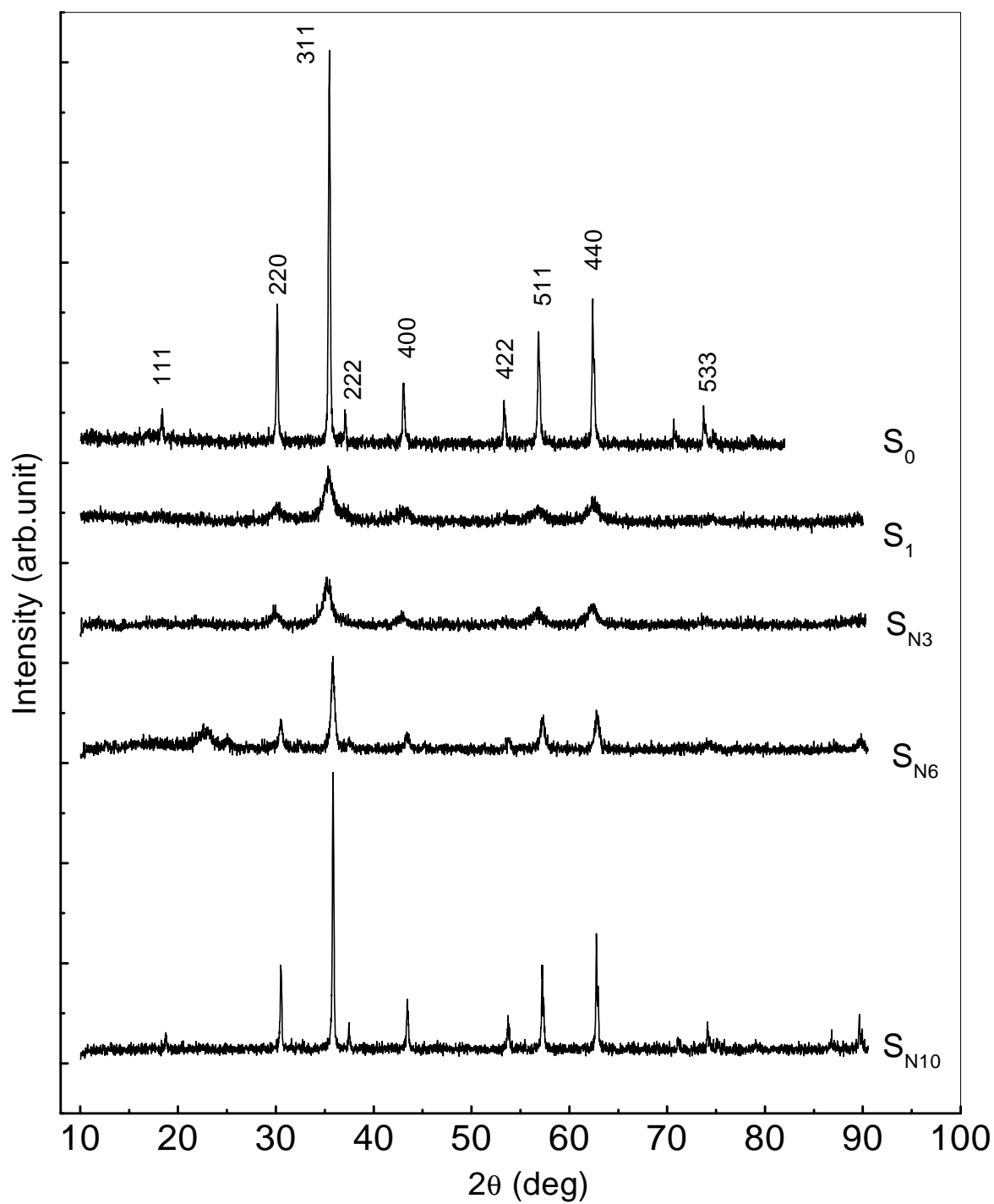
Fig.1 XRD spectra for Co$_{0.2}$Zn$_{0.8}$Fe$_2$O$_4$ spinel oxide heated at
different temperatures. For the notation S$_0$, S$_{1}$, S$_{N3}$, S$_{N6}$ and
S$_{N10}$ see in the text.\\
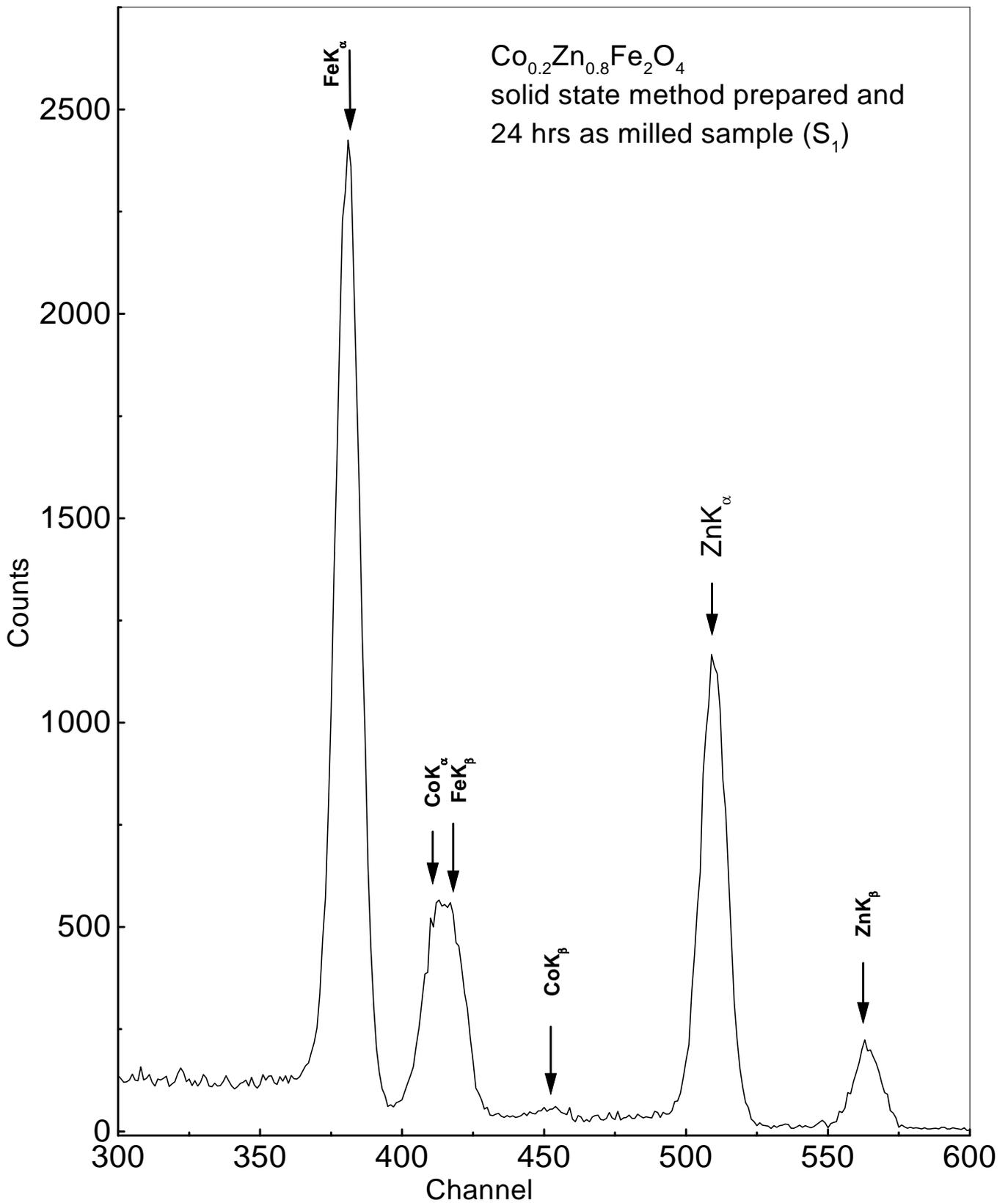
Fig.2  X-ray Fluorescence Spectrum for the as milled sample. the arrow
indicates the corresponding metals characteristics K$_{\alpha}$ and
K$_{\beta}$ lines.\\
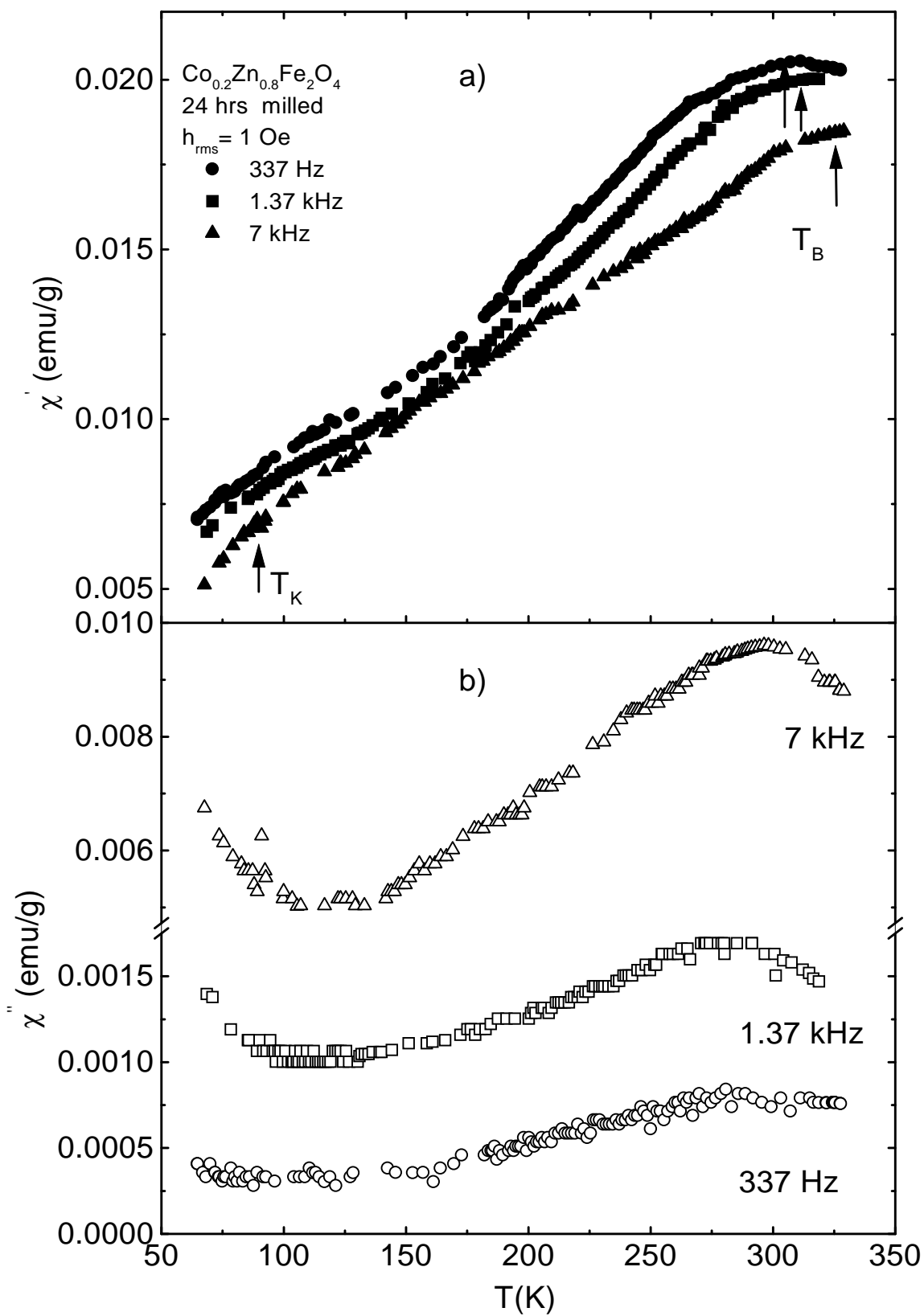
Fig.3 Ac susceptibity $\chi^\prime$ (in a) and $\chi^{\prime\prime}$ (in b)
data for as milled sample (S$_{1}$). T$_B$ and T$_K$ represent blocking and spin
canting temperature, respectively.\\
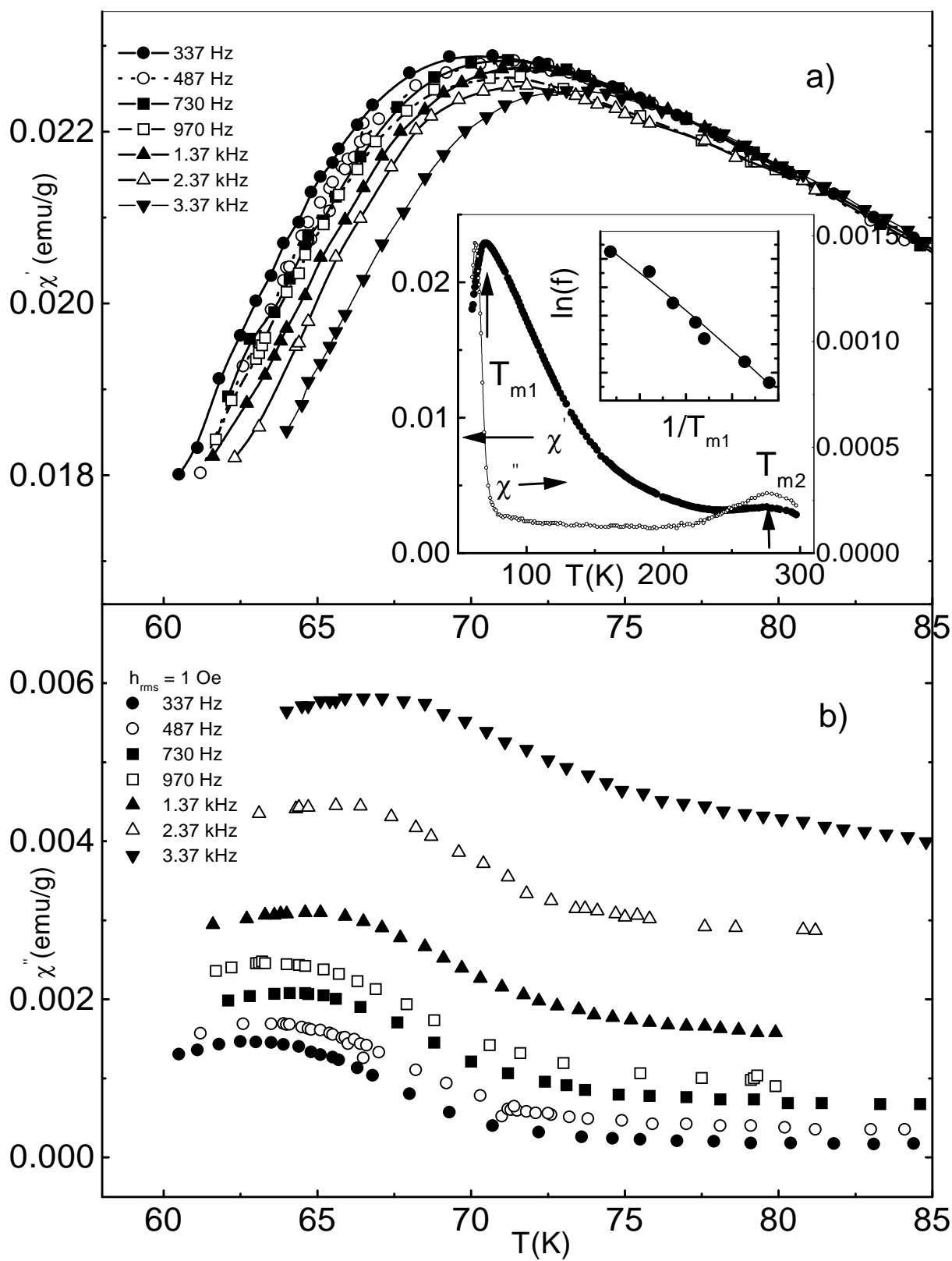
Fig.4 Ac susceptibity $\chi^\prime$ (in a) and $\chi^{\prime\prime}$ (in b)
for the 1000$^0$C-6 hour heat treated sample (S$_{N10}$). Fig.a inset shows ac
susceptibility data at 1 Oe ac field, f = 337 Hz and ln f vs 1/T$_{m1}$ plot
for the same sample.\\
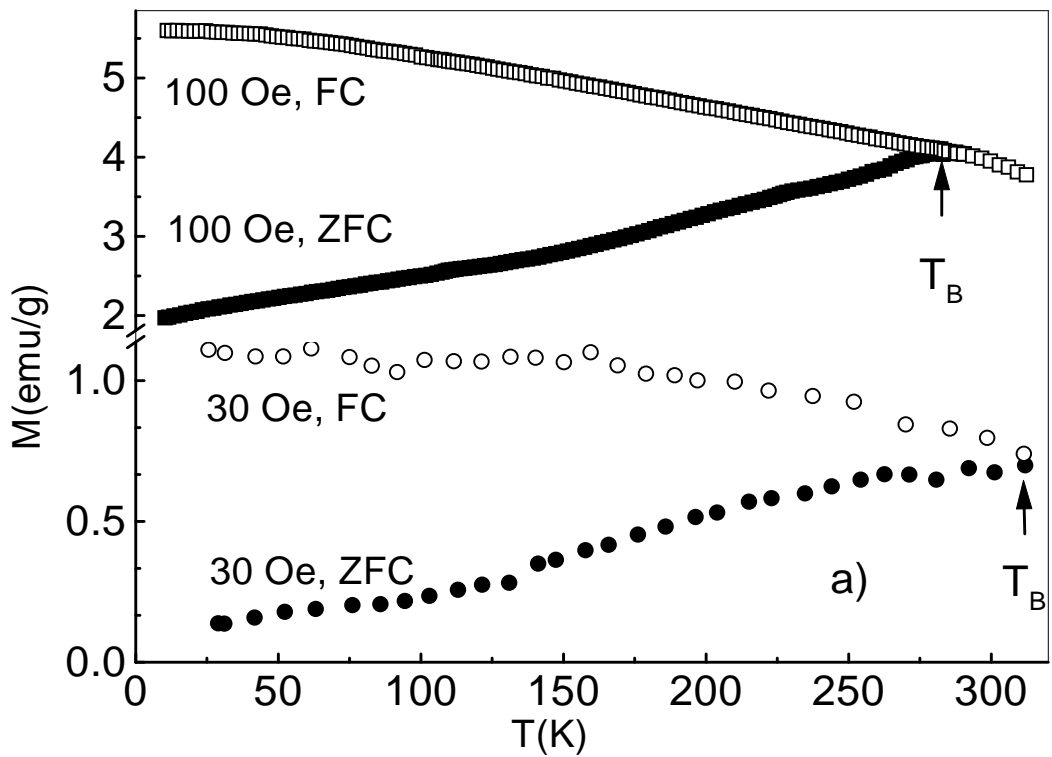
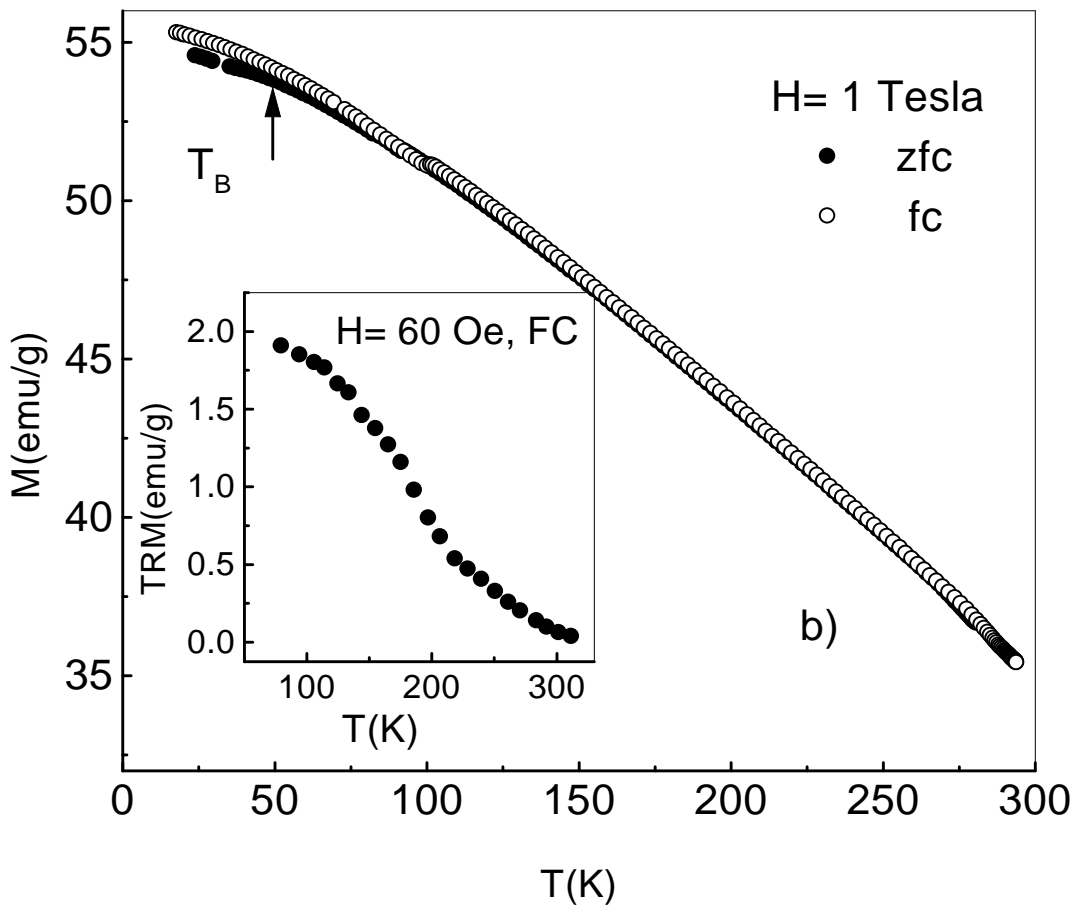
Fig.5 Dc magnetization vs temperature for 24 hrs milled 
Co$_{0.2}$Zn$_{0.8}$Fe$_2$O$_4$ sample. Fig. b inset shows the temperature
dependence of field cooled (FC) (cooled in a field of 60 Oe) thermoremanent 
magnetization (TRM).\\
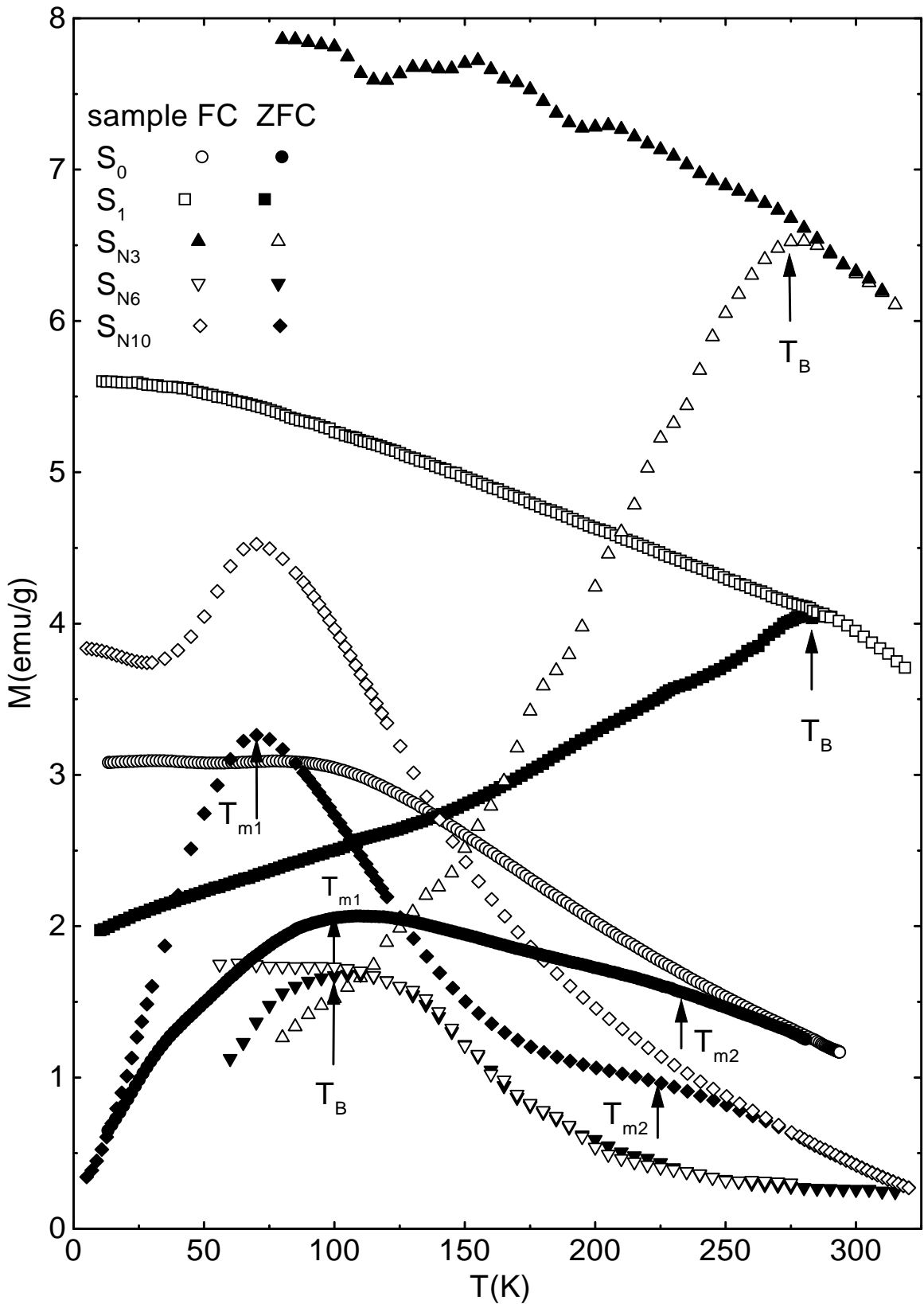
Fig.6 Temperature dependence of dc magnetization data for different heat
treated samples at 100 Oe field.\\
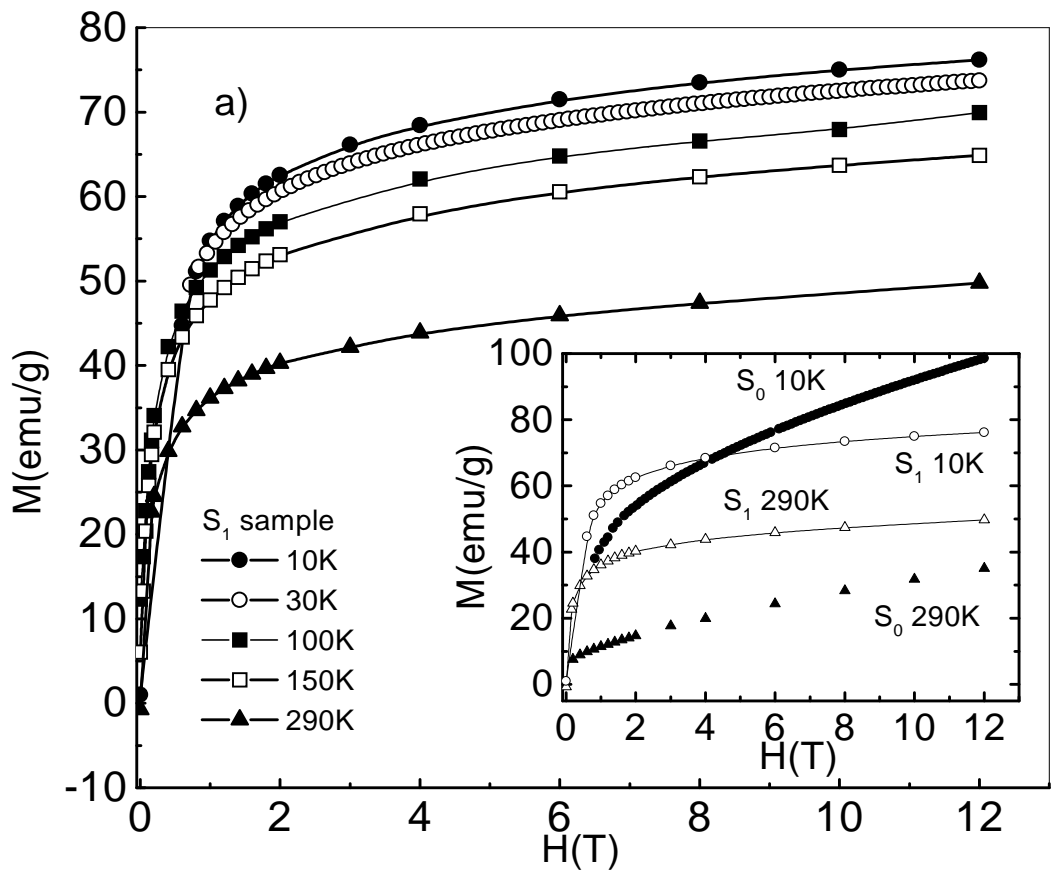
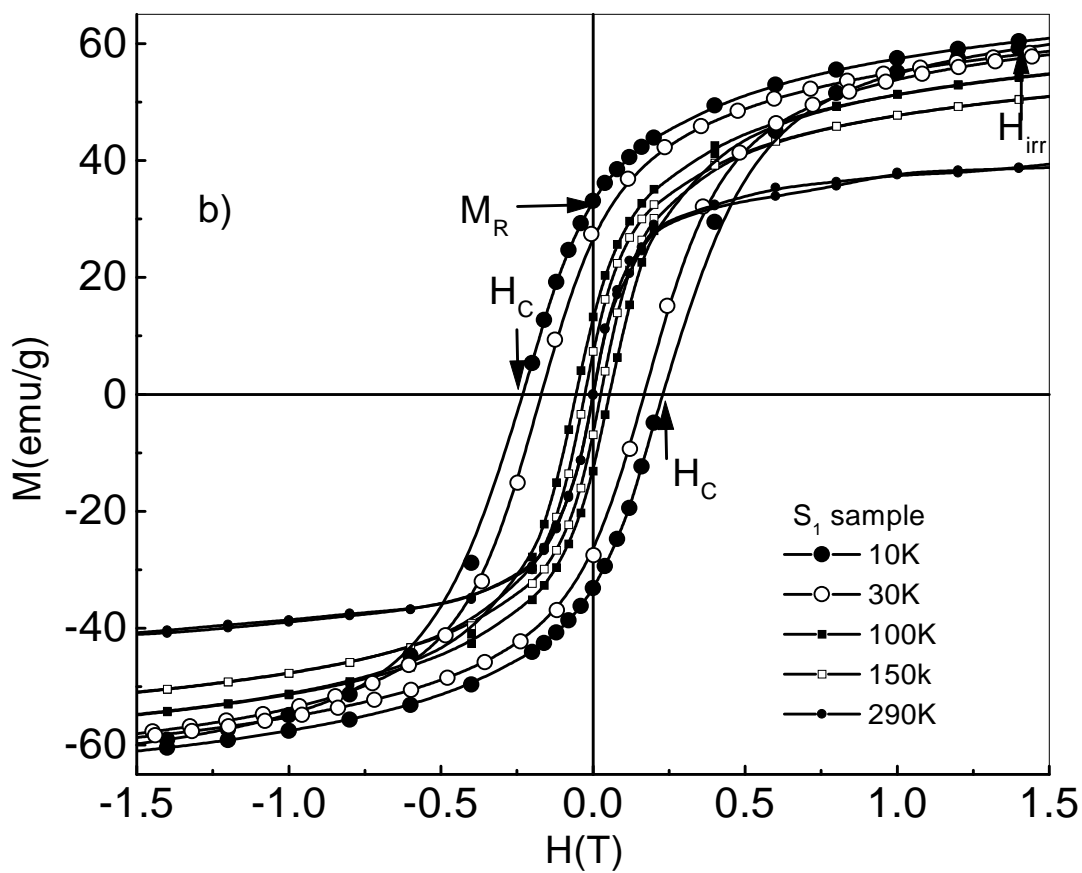
Fig.7 a) M vs H plot at different temperatures for S$_{1}$ sample and inset Fig.
compare the 10K and 290K data of S$_0$ and S$_{1}$ samples. b) Hysteresis loopes
at different temperatures for S$_{1}$ sample. For M$_R$, H$_C$ and H$_{irr}$ see
text.\\
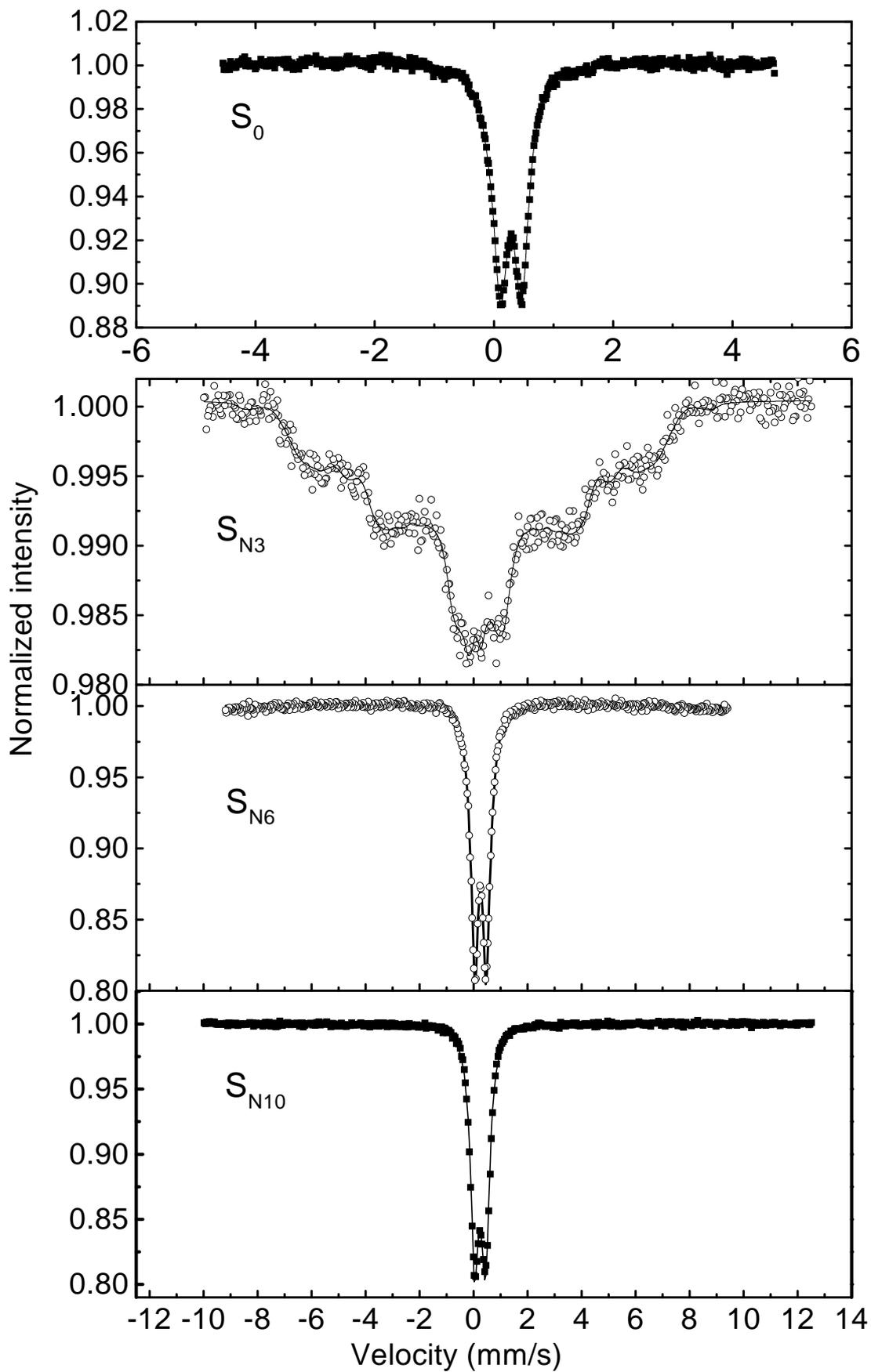
Fig.8  Room temperature M\"{o}ssbauer spectra recorded in absence of field for
as prepared bulk (S$_0$) and heat treated as milled $_{1}$, S$_{N3}$,
S$_{N6}$
and S$_{N10}$ sample.\\
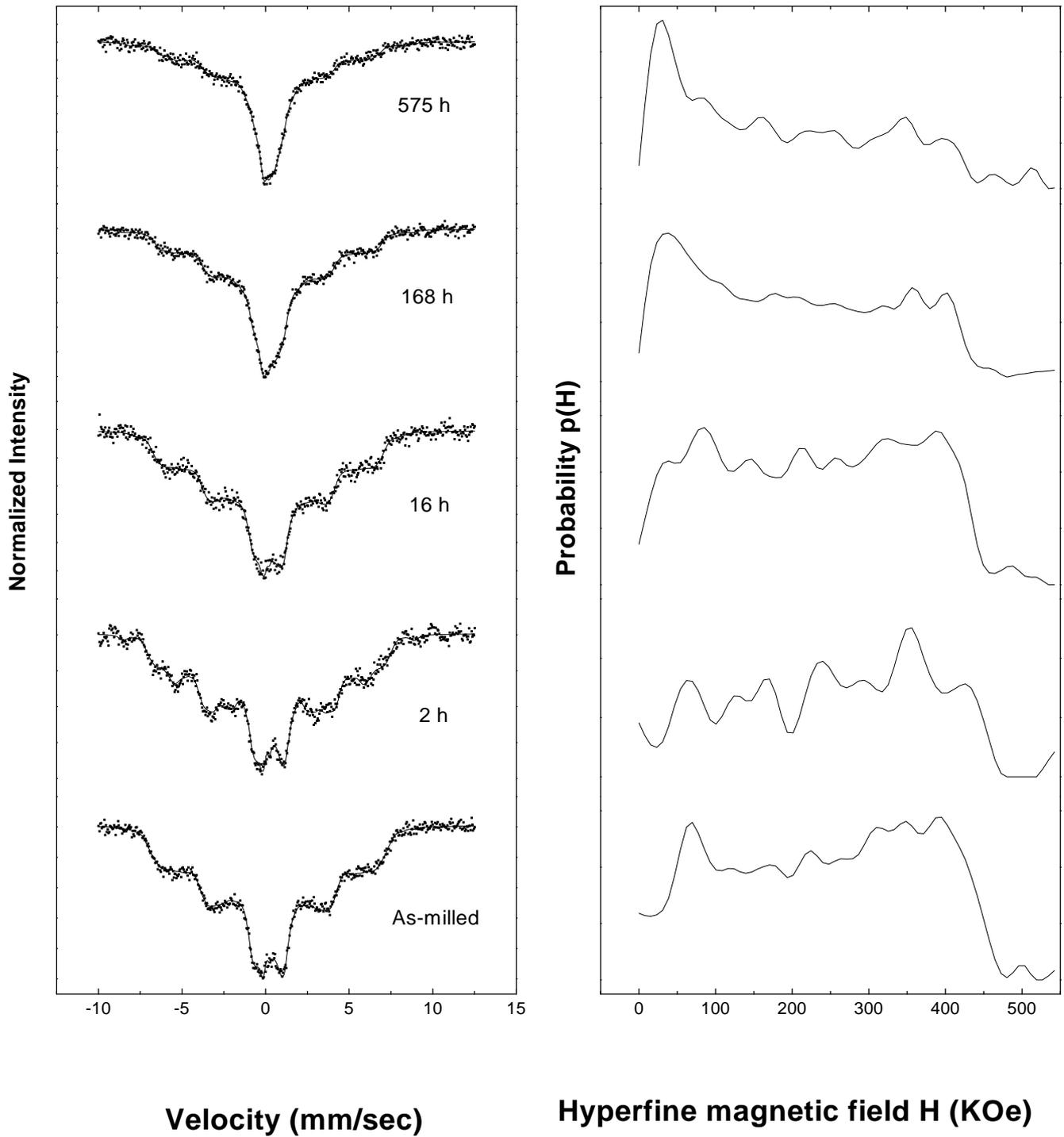
Fig.9 Room temperature M\"{o}ssbauer spectra for heat treated 
Co$_{0.2}$Zn$_{0.8}$Fe$_2$O$_4$ sample at 300$^0$C for different time (left)
and corresponding hyperfine field distribution (right).\\
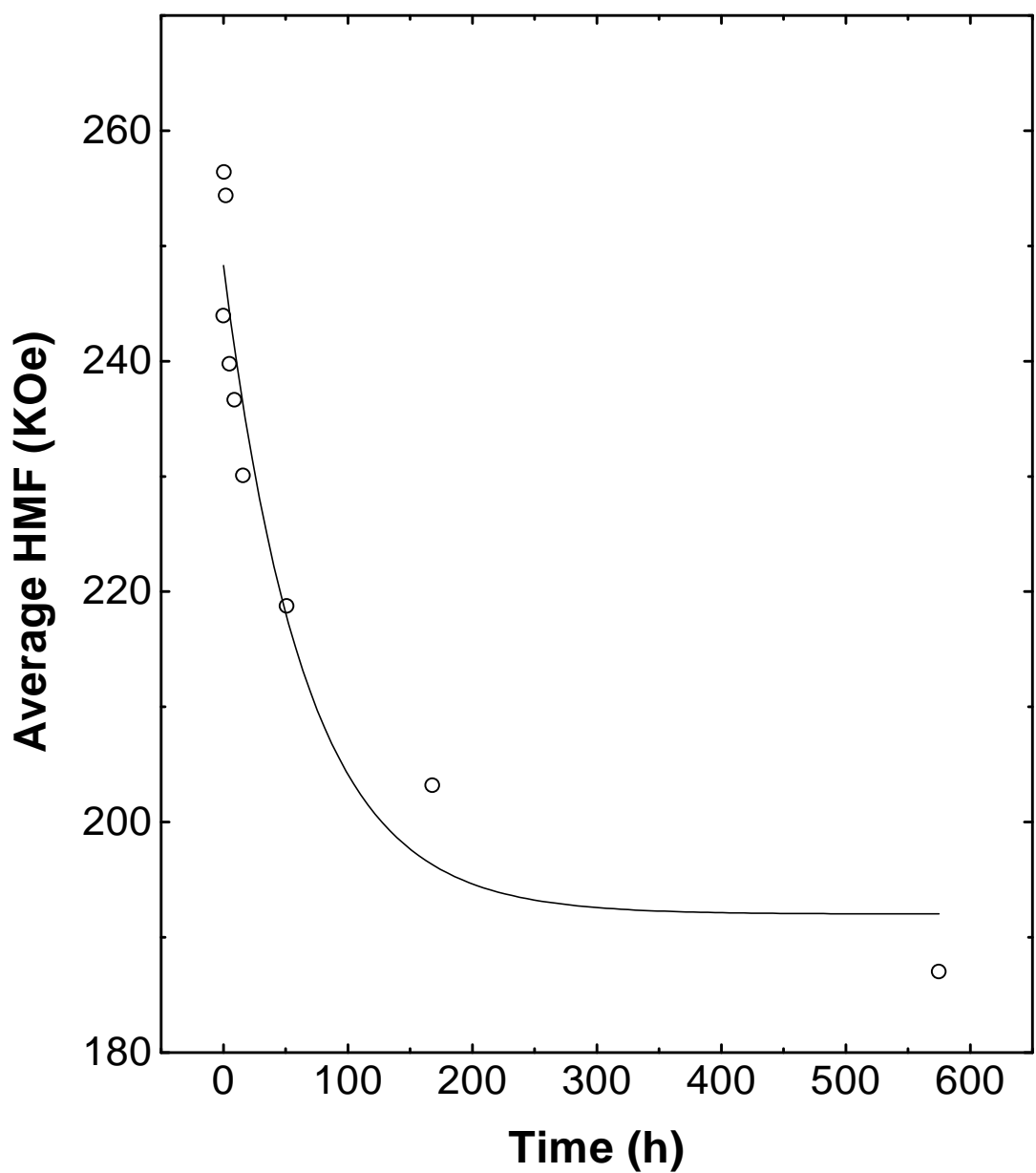
Fig.10  Annealing time (at 300$^0$C) dependence of average hyperfine field
value for as milled Co$_{0.2}$Zn$_{0.8}$Fe$_2$O$_4$ sample. 
\end{document}